\documentclass{article}

\usepackage{arxiv}

\usepackage[utf8]{inputenc} 
\usepackage[T1]{fontenc}    
\usepackage{hyperref}       
\usepackage{url}            
\usepackage{booktabs}       
\usepackage{amsfonts}       
\usepackage{nicefrac}       
\usepackage{microtype}      
\usepackage{lipsum}
\usepackage{graphicx}
\usepackage{textgreek}
\usepackage{amsmath}
\usepackage{adjustbox}
\usepackage{amssymb}
\usepackage{latexsym}
\usepackage{subcaption}
\usepackage{siunitx}  
\graphicspath{ {./images/} }

\title{U-net based prediction of cerebrospinal fluid distribution and ventricular reflux grading}

\author{
 Melanie Rieff$^{*,1}$, Fabian Holzberger$^1$, Oksana Lapina$^3$, Geir Ringstad$^{3,4,5,6}$, Lars Magnus Valnes$^2$\\
 \textbf{Bogna Warsza$^3$, Per Kristian Eide$^{2,5,6}$, Kent-Andr\'e Mardal$^{5,7,8}$, Barbara Wohlmuth$^1$}\\
 \\
  $^1$Department of Mathematics, School of
Computation, Information, and Technology \\
Technical University of Munich, Garching,
Germany \\
   \\
 $^2$Department of Neurosurgery, Oslo University, Hospital Rikshospitalet, Oslo, Norway \\
 \\
 $^3$Department of Radiology, Oslo University
Hospital Rikshospitalet, Oslo, Norway \\
\\
$^4$Department of Geriatrics and Internal
Medicine, Sorlandet Hospital, Arendal,
Norway \\
\\
$^5$KG Jebsen Centre for Brain Fluid Research,
Institute of Clinical Medicine, Faculty of
Medicine,\\ University of Oslo, Oslo, Norway \\
\\
$^6$Institute of Clinical Medicine, University of
Oslo, Oslo, Norway \\
\\
$^7$Department of Mathematics, University of
Oslo, Oslo, Norway \\
\\
$^8$Department of Numerical Analysis and
Scientific Computing, Simula Research
Laboratory, Oslo, Norway \\
\\
$^*$Corresponding author. E-mail \texttt{mrieff@student.ethz.ch}.
}

\begin{document}
\maketitle
\begin{abstract}
Previous work indicates evidence that cerebrospinal fluid (CSF) plays a crucial role in brain waste clearance processes, and that altered flow patterns are associated with various diseases of the central nervous system. In this study, we investigate the potential of deep learning to predict the distribution in human brain of a gadolinium-based CSF contrast agent (tracer) administered intrathecal. For this, T1-weighted magnetic resonance imaging (MRI) scans taken at multiple time points before and after injection were utilized. We propose a U-net-based supervised learning model to predict pixel-wise signal increase at its peak after 24 hours. Performance is evaluated based on different tracer distribution stages provided during training, including predictions from baseline scans taken before injection. Our findings show that training with imaging data from only the first two hours post-injection yields tracer flow predictions comparable to models trained with additional later-stage scans. Validation against ventricular reflux gradings from neuroradiologists confirmed alignment with expert evaluations. These results demonstrate that deep learning-based methods for CSF flow prediction deserve more attention, as minimizing MR imaging without compromising clinical analysis could enhance efficiency, improve patient well-being, and lower healthcare costs.
\end{abstract}


\section{Introduction}

The cerebrospinal fluid (CSF) flow and its role in solute transport and clearance has received a lot of attention since the introduction of the glymphatic system in 2012~\cite{Iliff2012, xie2013sleep}. In particular, the CSF and its interaction with the extra-cellular environment of the brain through perivascular spaces (PVSs) may be crucial for the clearance of metabolic waste, which is known to accumulate in Alzheimer's and Parkinson's diseases~\cite{rasmussen2018glymphatic}, to name a few. However, the CSF dynamics is complex as it is dictated by not only cardiac pulsations, but respiration~\cite{Dreha2015, vinje2019respiratory}, vasomotion~\cite{goodman2020vasomotor, kedarasetti2022arterial}, sleep waves~\cite{fultz2019coupled, bojarskaite2023sleep} as well as the circadian rhythm~\cite{hablitz2020circadian}.  Hence, the governing mechanisms are only partially known, and modeling attempts have as such only partially explained the tracer distribution in humans~\cite{Valnes2020, Vinje2023}. 
Furthermore, neuroanatomic characteristics can vary significantly between individuals, and differences in tracer distributions may be driven mostly by geometry difference. 
Given the partially unknown physical processes, it is interesting to investigate whether purely data-driven approaches can predict future tracer distributions and what types of data yield the best predictions.

Several novel imaging methods have been proposed for evaluating glymphatic function, including DTI-ALPS~\cite{taoka2017evaluation} and  imaging of enlarged PVSs~\cite{yamamoto2024perivascular}, which focus on structural features. Furthermore, flow imaging for cardiac or respiratory pulsations is available with phase-contrast MR~\cite{dreha2015inspiration, liu2024cardiac}, both in 2D slices and in 3D~\cite{vikner2024csf}.  These techniques, in particular, have been shown to be useful for the ventricular system and the cranio-cervical junction in conditions such as normal pressure hydrocephalus (NPH)~\cite{bradley2015csf, owashi2024phase} and the Chiari malformation~\cite{bhadelia2023cerebrospinal}, where the abnormal CSF pulsations at cardiac frequencies are used in diagnostics. However, it is not clear that such imaging is useful for solute transport and clearance that may be governed by slower processes. As such, one new and promising method is exploiting intrathecal MR contrast to visualize the slow transport of gadobutrol in the CSF and the extracellular space of the brain~\cite{iliff2013brain, Ringstad2017}. A main advantage of the technique is that imaging illustrates the solute transport on the long time-scale of days and it has been demonstrated that there is great variation among patients~\cite{Eide2021_2}, and that the transport is reduced in sleep-deprived individuals~\cite{Eide2021_3}. A major drawback of the procedure is that it is time-consuming, as patients undergo multiple imaging sessions over several days. Furthermore, only a few centers worldwide currently perform intrathecal MR contrast investigations, as the methodology is off-label (although it is occasionally used for assessing CSF leaks in spontaneous intracranial hypotension~\cite{patel2020ct}). \par
Modeling and/or data-driven approaches could significantly reduce patient scanning time if they provide accurate predictions. Furthermore, since the procedure is off-label, one may question whether tracer distributions can be learned solely from anatomy, as seen in a standard T1-weighted image. \par
In this study, we adapt a modification of U-net \cite{Ronneberger2015}, a specific type of convolutional neural network (CNN),  to tracer analysis and hereby evaluate the ability of deep learning methods to predict CSF distribution patterns of 2D MR imaging data. In addition to evaluating the image similarity of predicted scans compared to real imaging data using evaluation metrics common in deep learning, we additionally use a clinically oriented reflux assessment scheme as a method to judge the model performance. This is significant because ventricular reflux assessment has been clinically useful in the preoperative diagnostic workup of idiopathic normal pressure hydrocephalus \cite{Eide2020, Eide2021_2}. The present findings showcase that even basic machine-learning techniques can be used to model complex physiological processes, and more specifically and uniquely in this context, demonstrate the use of deep learning techniques in CSF distribution prediction. In the future, it is hoped that this approach could avoid the need for repeated MRIs, thereby reducing medical costs, clinical resource consumption, and patient burden.

\section{Related work}
Previous research on cerebrospinal fluid analysis has employed various mathematical and computational modeling approaches to understand its dynamics and interplay with several neurological conditions. CNNs, for example, were utilized numerous times to predict AD from brain MRI data, though almost exclusively not contrast-enhanced. Pathological molecular hallmarks of AD, such as amyloid-β-42 and tau proteins, were quantified using an ensemble-based model\cite{Popuri2020}. Schweizer et al. examined the potential to use CNNs for classification of individual cells that CSF contains\cite{Schweizer2023}, and real-time CSF flow velocities were estimated using inflow effect-mediated signal increase of in vivo functional MRI\cite{Diorio2024}. Four-dimensional flow of CSF has been assessed with various patient cohorts, fluid markers, and regions of interest\cite{Rivera2024}, although the limitations of this imaging approach due to low velocities associated with CSF motion and large T1 relaxation values were continuously faced. \par 
While well-established mathematical approaches such as mesh-based finite element methods and partial differential equation-constrained optimization were used to directly model diffusion of CSF in the human brain\cite{Valnes2020}, they face some drawbacks such as high computational cost and the need of high-quality meshes to simulate complex geometries. Quite recently, physics-informed neural networks (PINNs) were introduced to integrate physical laws into neural networks for various applications, including heat transfer and diffusion. Zapf et al.\cite{Zapf2022} highlighted the potential of PINNs in solving inverse problems governed by partial differential constrained optimization, which were extended to estimate the apparent diffusion coefficient of cerebrospinal fluid tracers in specific regions of interest in the brain (subcortical white matter subregions and full brain parenchyma) and experimentally validated with MRI data from sleep-deprived patients\cite{Vinje2023}. Their findings suggest that diffusion alone may not be sufficient to model molecule distribution in the brain, supporting the glymphatic model's perspective on molecular transport. In the domain of computational fluid dynamics, physiological boundary conditions improved the simulations of intracranial pressures and flow velocities through the cerebral aqueduct and spinal SAS\cite{Vandenbulcke2022}. \par
In this study, we explore the application of deep learning techniques to predict CSF tracer distribution; an approach that has not been previously applied to this problem. By doing so, we aim to demonstrate the potential of a simple data-driven machine-learning framework for modeling complex physiological processes, without explicitly requiring anatomical modeling of CSF production, circulation, or volume properties.

\section{Materials and Methods}\label{sec2}
\subsection{Study population}
This work was a single-center retrospective study that utilized MRI data from a cohort of 136 patients with different neurological conditions. All participants provided informed consent prior to their involvement in the study. Table \ref{table:patients} shows the distribution of neuropathological findings for all of the analyzed data. This distribution of conditions reflects the typical case mix observed in our clinical practice for individuals requiring CSF assessment, underscoring its representativeness for this specific patient population at our center. For training and testing our model, we randomly split the \text{N\textsubscript{total}} = 136 cases in \text{N\textsubscript{train}} = 105 and \text{N\textsubscript{test}} = 31.

\begin{table*}[h]%
\centering %
\caption{Patient material}%
\begin{tabular*}{\textwidth}{@{\extracolsep\fill}lcccc@{\extracolsep\fill}}
\toprule
 & \textbf{N\textsubscript{total}}  & \textbf{Age} & \textbf{Sex (Female/Male)}& \textbf{Body mass index \si{kg.m^{-2}} } \\
\midrule
Total cohort & 136  & 48.0 $\pm$ 18.2 & 85/51 & 27.7 $\pm$ 5.1     \\
References & 25 & 35.9 $\pm$ 10.4 & 21/4 & 27.9 $\pm$ 5.4   \\
Arachnoid cyst (AC) & 17 &48.1 $\pm$ 16.9 & 9/8 & 26.7 $\pm$ 2.8  \\
Pineal cyst (PC)  & 14 & 37.1 $\pm$ 13.0 & 13/1 & 28.1 $\pm$ 4.7  \\
Idiopathic intracranial hypertension (IIH) & 16 & 34.6 $\pm$ 11.8 & 14/2 & 31.7 $\pm$ 5.1 \\
Spontaneous intracranial hypotension (SIH) & 15  & 48.1 $\pm$ 14.7 & 11/4 & 25.7 $\pm$ 6.2   \\
Idiopathic normal pressure hydrocephalus (iNPH) & 33  & 70.8 $\pm$ 6.5 & 9/24 & 27.0 $\pm$ 4.2   \\
Hydrocephalus (HC) & 16  & 42.6 $\pm$ 14.5 & 8/8 & 27.3 $\pm$ 6.0     \\
\bottomrule
\end{tabular*}
\label{table:patients}
\end{table*}

\subsection{Injection protocol, MRI hardware, and data preprocessing}
To evaluate the enrichment of a CSF tracer in the human brain, the MRI contrast agent gadobutrol was utilized. An interventional neuroradiologist administered an intrathecal injection of gadobutrol at a dosage of \SI{0.5}{\milli\mol} (\SI{0.5}{\milli\liter} of \SI{1.0}{\milli\mol\per\milli\liter} gadobutrol; Gadovist, Bayer Pharma AG, Germany). T1-weighted MRI scans were performed before and at various intervals post-injection. For practical reasons, MRI examinations could not be conducted at the exact same times for each study participant, but there is consistency grouping the scans into the following time intervals: 0 hours (before injection, baseline), 1-2 hours, 3–5 hours, 5-6 hours, 6-10 hours, 24 hours, 48 hours, and 29 days (full clearance, control scan). \par
We used a 3 Tesla Philips Ingenia MRI scanner (Philips Medical Systems) with equal imaging protocol settings at all time points to acquire sagittal 3D T1‐weighted scans. The imaging parameters were: repetition time = ‘shortest’ (typically \SI{5.1}{\milli\second}), echo time = ‘shortest’ (typically \SI{2.3}{\milli\second}), flip angle=\SI{8}{\degree}, \SI{256}{\milli\meter}\texttimes\SI{256}{\milli\meter} field of view and matrix of size 256\texttimes256 (reconstructed 512\texttimes512). The slices were reconstructed with \SI{1}{\milli\meter} thickness that were automatically reconstructed to 368 slices with \SI{0.5}{\milli\meter} distance; total duration of each image acquisition was \SI{6}{\minute} and \SI{29}{\second}. To secure consistency and reproducibility of the MRI slice placement and orientation, slice orientation of image stacks was defined using an automated anatomy recognition protocol based on landmark detection in MRI data (SmartExamTM, Philips Medical Systems) for every time point.
The images for each patient were registered/aligned using FreeSurfer software (version 6.0) (\url{http://surfer.nmr.mgh.harvard.edu/}) and re-sampled to the FreeSurfer standard of \SI{256}{\textsuperscript{3}} voxels with voxel size of  \SI{1}{\milli\meter}\texttimes\SI{1}{\milli\meter}\texttimes\SI{1}{\milli\meter}\cite{Eide2021_3, Eide2021_2, Ringstad2018}. The images was normalized by dividing the signal values with the mean signal value of a reference volume within the intraconal orbital fat\cite{Eide2021_3}. Then, images were each centered and the sagittal and axial slices were exported as 256\texttimes256 matrices. In this study, intensity scaling between 0 and 1 was performed independently for each 2D slice to normalize intensity variations within slices, ensuring consistent signal distribution for model training and analysis; however, as this approach may inadvertently introduce artifacts in non-enhanced tissue intensities, and alternative global normalization strategies will be explored in future work. For predicting tracer distribution after 24 hours, we fed our model with different information during training and compared the results. \par
Importantly, for each patient and time point, we used one 2D scan slice in the axial plane and one in the sagittal plane as in-/output to the network, which we refer to as an image pair in the following. Predicted sagittal and transverse slices provide complementary views of the ventricular system, and both planes are valuable for grading ventricular reflux. The sagittal slice is particularly well-suited for assessing CSF tracer enhancement in the third ventricle relative to the suprasellar cistern below, determining whether the enhancement is equal or differs. In contrast, the axial slice is ideal for evaluating enhancement within the ventricles and the corresponding level of the subarachnoid space outside the cerebral convexities. In principle and given sufficient computational resources, our approach can be extended to volumetric data.

\subsection{Distribution pattern}
\begin{figure}
  \centering
  \includegraphics[page=1,width=.7\textwidth]{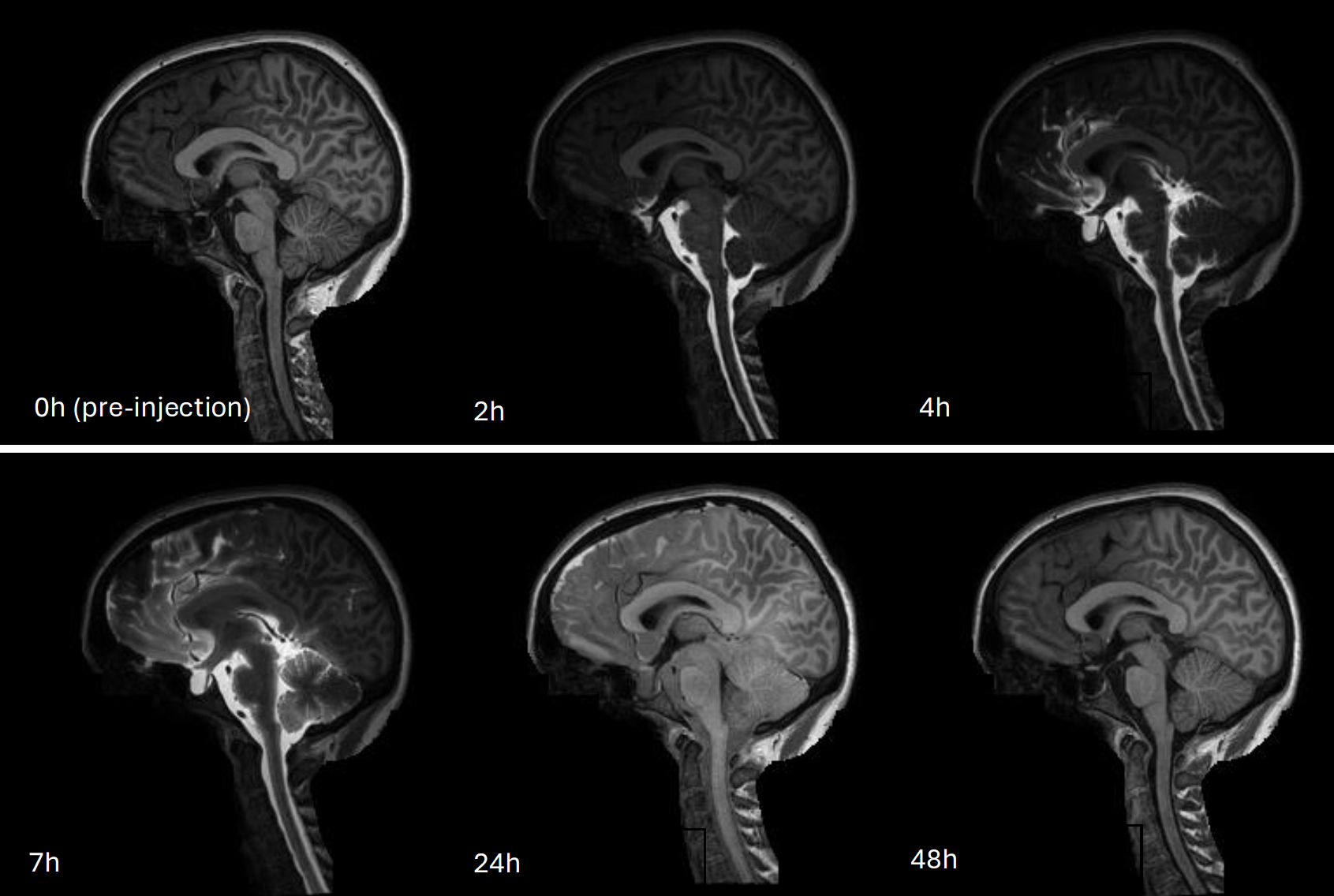}
  \caption{Defaced sagittal slices from CSF tracer–enhanced MRI of a sample patient. After 24 hours, the tracer has enriched the CSF spaces around the entire brain and is completely cleared after four weeks.}
  \label{fig:sample_scans}
\end{figure}

A series of T1-weighted images of the patient taken after intrathecal tracer administration, exemplified in Fig. \ref{fig:sample_scans}, show the tracer transport and elimination mechanisms. Following the injection of gadobutrol into the spinal canal at the lumbar region, the tracer ascends through the spinal canal, and typically reaches the subarachnoid space (SAS) at the level of the foramen magnum\cite{Liu2022}. Significant spinal resorption means that only about \SI{25}{\percent} of the injected tracer reaches intracranial space\cite{Vinje2023}. Once in the cisterna magma, the tracer rapidly moves to subarachnoid cisterns and then distributes along major arteries within the cerebral fissures. To evaluate the enrichment of a CSF tracer in the human brain, the MRI contrast agent gadobutrol was utilized as CSF tracer, showing significant accumulation in areas adjacent to the major arteries\cite{Vinje2023}. MRI scans performed 24 hours after injection reveal diffuse tracer enhancement around the brain, with signal enhancement peaking typically at 24 hours after injection\cite{Ringstad2017}. Gadobutrol clearance is driven by resorption initially from the spinal canal, and peak gadobutrol concentration in the blood occurs around five to six hours after injection\cite{Eide2021_1, Hovd2022, Shah2023, Melin2023}. At the four-week post-injection mark, MRI scans reveal no evidence of contrast agent accumulation in the brain. Still, the exact underlying mechanisms of tracer distribution and clearance remain not fully understood \cite{Plog2018}.

\subsection{Model architecture}
\begin{figure}[ht]
  \centering
  \includegraphics[clip, , trim=0.5cm 17cm 0.5cm 4cm ,width=.8\textwidth]{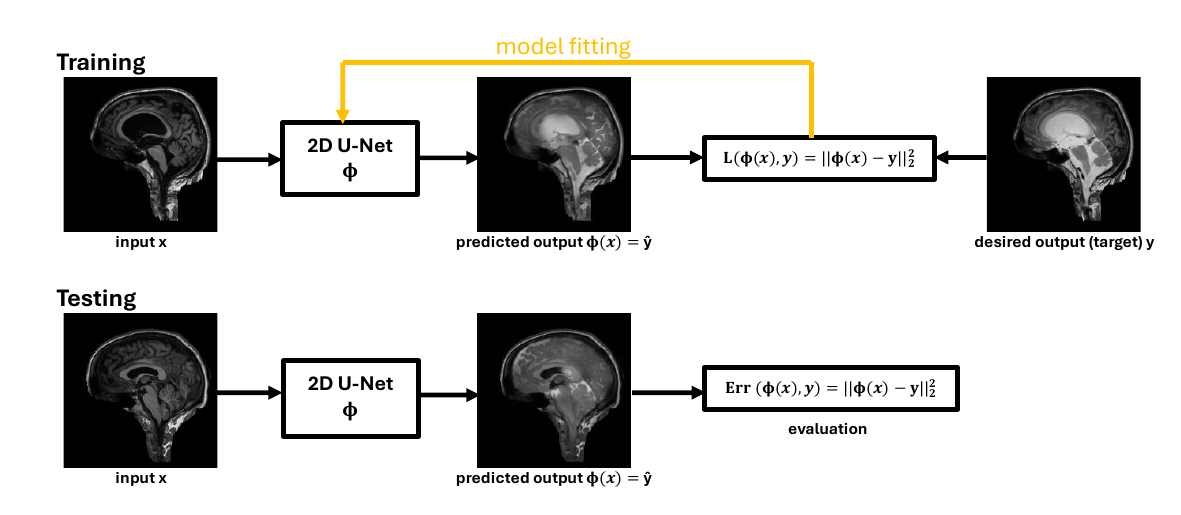}
  \caption{Exemplary illustration of the training and testing steps. Here, the network was trained by minimizing the $\ell_2$-loss of the reconstructed sagittal images to the real images and final model evaluation made use of the mean of all squared errors among the testing data.}
  \label{fig:training_process}
\end{figure}

Inference in neural networks can be described by the equation $ \hat{y}= f(x; W)$, where $x$ is the input, $W$ are weights of the network, and $\hat{y}$ is the output, i.e., the prediction. Training a neural network corresponds to finding the solution $\hat{W}$ of the loss minimization problem
\begin{equation*}
    \hat{W}  = \arg\min_{W} L(W, \mathcal{D}) =  \arg\min_{W} \frac{1}{n} \sum_{i=1}^n \ell(f_W(x_i), y_i),
\end{equation*}
where $\ell$ is a given loss function and $\mathcal{D} = \{(x_1, y_1), (x_2, y_2), \dots, (x_n, y_n)\}$ is the training data. There exists a variety of loss functions.
Here, we use a regression model and compare the two commonly chosen loss functions: the $\ell_2$-loss defined as $\ell_2(\hat{y},y) = (\hat{y}-y)^2$, and the $\ell_1$-loss, defined as $\ell_1(\hat{y},y) = |\hat{y}-y|$. Experimenting with different loss functions is a crucial part of designing a deep learning model, because the choice of the loss function significantly impacts the model's performance, behavior, and convergence during training, which has been shown a number of times \cite{Shan2023}. Figure \ref{fig:training_process} depicts a simplified illustration of the training process.

In this study, we used a U-net architecture for the prediction of the tracer distribution. Introduced by Olaf Ronneberger et al. in 2015\cite{Ronneberger2015}, U-net is a model originally proposed for 2D or 3D image segmentation, particularly in biomedical applications\cite{Cai2022, Siddique2021}. Thereafter, it served as a blueprint for a plethora of variations that have so far been proposed and used in several application areas\cite{Wang2023, Reza2022}, which all share a common structure inspired by an encoder-decoder architecture comprising two primary paths: the contracting (or compressive) path and the expansive (or up-sampling) path. The contracting path functions similarly to a convolutional neural network (CNN), where images are down-sampled at first using a convolutional kernel followed by an activation function. Conversely, the expansive path up-samples the compressed representation to reconstruct the image. In addition, skip connections link the corresponding layers of the contracting and expansive paths, helping to preserve spatial information that is typically lost during the down-sampling process\cite{wilm2024}. In contrast to other supervised deep learning methods such as CNNs and fully convolutional neural networks (FCNs), U-nets have a more complicated architecture, yet have been shown to perform better, particularly on medical imaging data and small data sets\cite{Wang2023}. Furthermore, U-Net-based models are well-equipped to handle small structures and objects within an image, a capability crucial for accurately predicting tracer fluid distribution. \par
\begin{figure}[ht]
  \centering
  \begin{subfigure}{\linewidth}
    \centering
  \includegraphics[page=1,width=.8\textwidth]{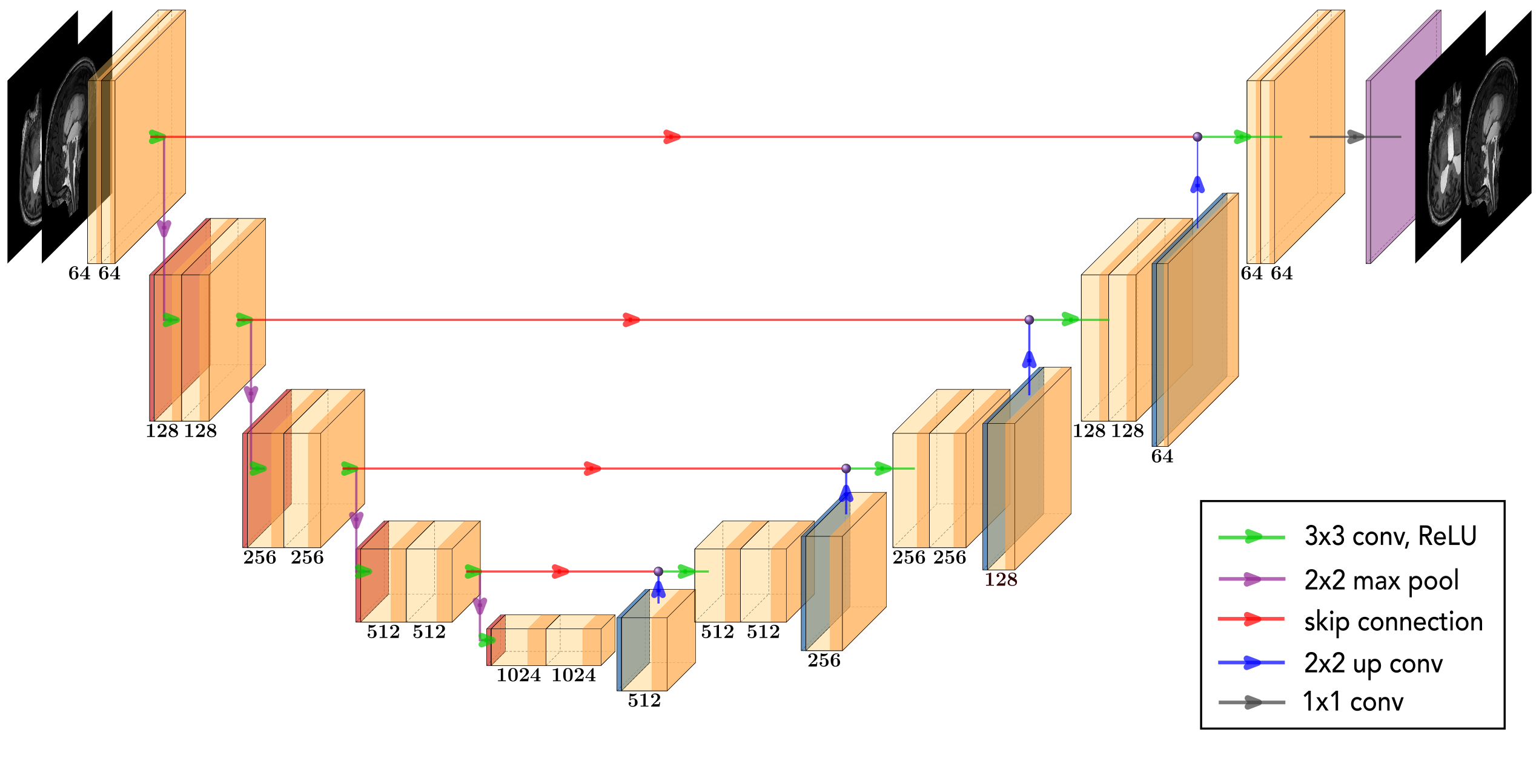}
    \caption{}
     \label{fig:UNET}
  \end{subfigure}
  \begin{subfigure}{\linewidth}
    \centering
    \includegraphics[clip, , trim=0.5cm 20cm 0.5cm 2.5cm ,width=.8\textwidth]{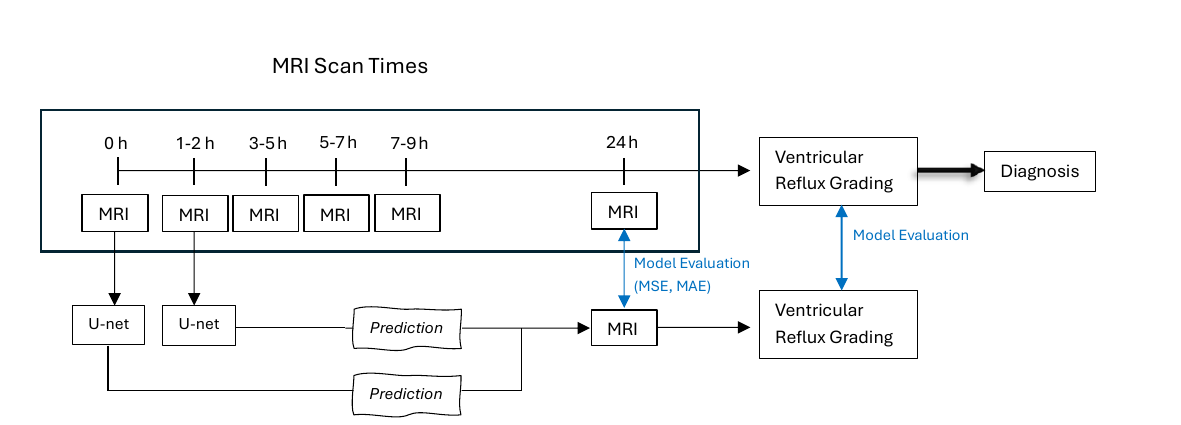}
    \caption{}
     \label{fig:processing_workflow}
  \end{subfigure}  
  \caption{\ref{fig:UNET}: Modified U-net architecture used in this work. Here, axial and sagittal MRI planes (one each) scanned before tracer injection are used to predict its distribution 24 hours after tracer injection. \ref{fig:processing_workflow}: Processing workflow from original MRI scans to U-Net model predictions, comparison with real MRI images, and final clinical diagnosis.}  
\end{figure}

For the implementation of our 2D U-Net architecture, we used PyTorch, a variable number of input channels, 64 features, 8 batch size, Adam optimizer, 200 or 250 epochs, \SI{1e-3}{} learning rate, 3×3 convolutional kernels followed by batch normalization and rectified linear unit as activation function, and 2×2 maximal pooling size. Firstly, we use a U-net as illustrated in Figure \ref{fig:UNET} with two input channels (for sagittal and axial scans, respectively) to use scans from baseline (before injection) as training input. With this, we address how well our model may accurately predict tracer distribution after 24 hours providing it with specific patient anatomy only. Changing the number of input channels allows the model to further use tracer-enriched scans. By comparing the model performance of varying training data, we aim to assess to what extent  different tracer distribution stages contribute to better prediction performance. Figure \ref{fig:processing_workflow} provides an overview of the data processing workflow, deep-learning model usage, and clinical interpretation.

\subsection{Ventricular reflux grading}
Traditionally, CSF diagnostics with MRI have primarily focused on anatomical abnormalities in the brain\cite{Ahmad2021}. To analyze cerebrospinal fluid dynamics, a grading scheme was introduced and since then used in clinical assessment of patients \cite{Eide2020}: Ventricular reflux refers to the retrograde movement of CSF from the SAS back into the cerebral ventricles and is measured on an integer scale from 0 to 4. Visually assessed reflux can be categorized into five grades based on the intensity and duration of tracer enrichment within the ventricles, which serves as a critical parameter in evaluating fluid dynamics and can be linked to the efficiency of CSF clearance\cite{Eide2021_2}:

\begin{itemize}
    \item Grade 0: No supra-aqueductal reflux.
    \item Grade 1: Sign of supra-aqueductal reflux.
    \item Grade 2: Transient enrichment of lateral ventricles.
    \item Grade 3: Lasting enrichment of lateral ventricles at 24 hours, not isointense with CSF subarachnoid.
    \item Grade 4: Lasting enrichment of lateral ventricles at 24 hours, isointense with CSF subarachnoid.
\end{itemize}
For detailed visual examples of these grades, readers are referred to the original grading studies conducted by Eide et al.\cite{Eide2021_2}
Reflux grades 3 and 4 have been shown associated with raised intracranial pressure pulsatility, a feature of shunt-responsive iNPH\cite{Eide2021_2, Eide2020}. Comparing MRI biomarkers of CSF tracer dynamics (including ventricular reflux grades) with anatomical MRI biomarkers of CSF space, markers of neurodegeneration and pulsatile intracranial pressure score showed reduced tracer clearance and higher grades of ventricular reflux (grades 3–4 in shunt-responsive idiopathic normal pressure hydrocephalus) \cite{Eide2020}. In this work, we use the ventricular reflux grading scheme as an additional method of model performance evaluation.

\section{Experimental Results}\label{sec3}
\subsection{Image prediction from baseline and tracer-enriched scans}
\begin{figure}[hbt!]
\centering
\begin{adjustbox}{minipage=\linewidth,scale=1}
\begin{subfigure}{.16\linewidth}
  \caption{}
  \includegraphics[width=\linewidth]
  {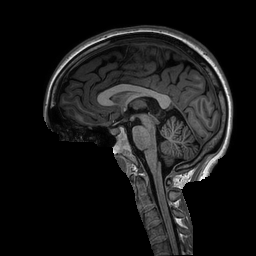}
  \label{inpt}
\end{subfigure}\hfill 
\begin{subfigure}{.16\linewidth}
\caption{}
  \includegraphics[width=\linewidth]{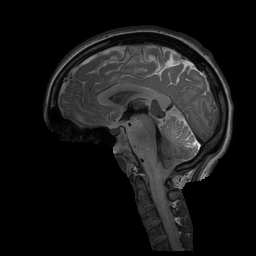}
  \label{tgt}
\end{subfigure}\hfill 
\begin{subfigure}{.16\linewidth}
\caption{}
  \includegraphics[width=\linewidth]{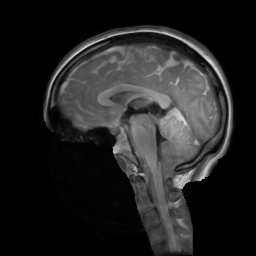}
  \label{mse}
\end{subfigure}\hfill 
\begin{subfigure}{.16\linewidth}
\caption{}
  \includegraphics[width=\linewidth]{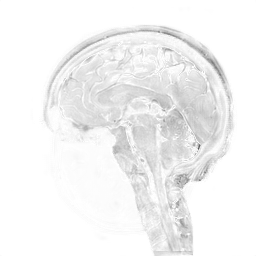}
  \label{mse_diff}
\end{subfigure}\hfill 
\begin{subfigure}{.16\linewidth}
\caption{}
  \includegraphics[width=\linewidth]{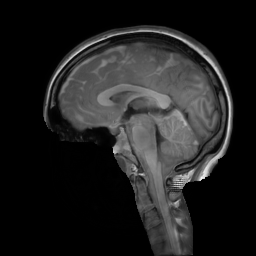}
  \label{mae}
\end{subfigure}
\begin{subfigure}{.16\linewidth}
\caption{}
  \includegraphics[width=\linewidth]{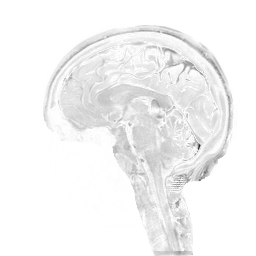}
  \label{mae_diff}
\end{subfigure}

\begin{subfigure}{.16\linewidth}
  \includegraphics[width=\linewidth]{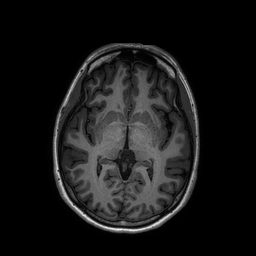}
\end{subfigure}\hfill 
\begin{subfigure}{.16\linewidth}
  \includegraphics[width=\linewidth]{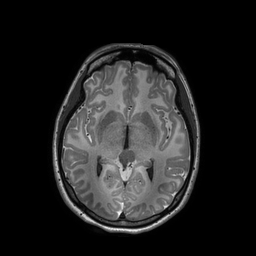}
\end{subfigure}\hfill 
\begin{subfigure}{.16\linewidth}
  \includegraphics[width=\linewidth]{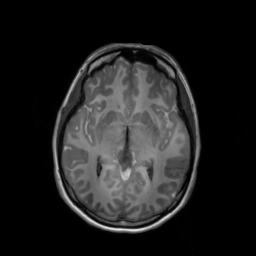}
\end{subfigure}\hfill 
\begin{subfigure}{.16\linewidth}
  \includegraphics[width=\linewidth]{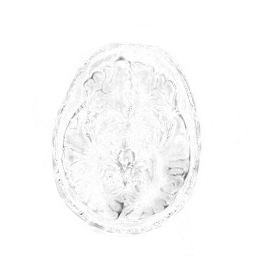}
\end{subfigure}\hfill 
\begin{subfigure}{.16\linewidth}
  \includegraphics[width=\linewidth]{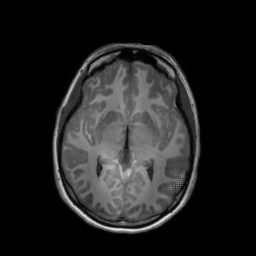}
\end{subfigure}
\begin{subfigure}{.16\linewidth}
  \includegraphics[width=\linewidth]{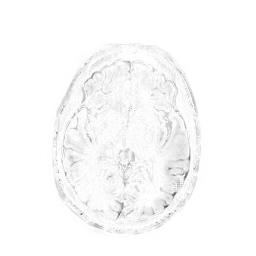}
\end{subfigure}
\end{adjustbox}
\caption{A sample test case (sagittal and axial plane) of gadobutrol distribution prediction based from baseline MRI scans (pre-injection). \ref{inpt}: real MR imaging taken before injection. \ref{tgt}: real MR imaging taken approximately 24 hours after intrathecal tracer injection. \ref{mse}: predicted tracer distribution 24 hours post-injection using  an $\ell_2$ loss function. \ref{mse_diff}: absolute difference between \ref{tgt} and \ref{mse}. \ref{mae}: predicted tracer distribution using an $\ell_1$loss function. \ref{mae_diff}: absolute difference between \ref{tgt} and \ref{mae}.}
\label{fig:output1}
\end{figure}

Mean squared error (MSE) $\frac{1}{n} \sum_{i=1}^n (\hat{y_i} - y_i)^2$ and mean absolute error (MAE) $\frac{1}{n} \sum_{i=1}^n |\hat{y_i} - y_i|$  of predicted $\hat{y_i}$ and real scans $y_i$, each of dimension 256×256×2 (axial and sagittal plane), were computed using PyTorch. We trained for 250 epochs and achieved \SI{7e-3}{} MSE and \SI{3.8e-2}{} MAE for the testing set and \SI{1e-3}{} MSE and \SI{1e-2}{} MAE for the training set, indicating overfitting of the model after 50 to 100 epochs. Visual inspection of predictions showed preservation of anatomical characteristics despite some blurriness. Figure \ref{fig:output1} presents a sample test case. A comparison between columns \ref{mse} and \ref{mae} with \ref{tgt} reveals that the model slightly underestimates tracer enrichment intensity, particularly in the parietal lobe. Furthermore, comparing columns \ref{mse_diff} and \ref{mae_diff} with \ref{tgt} demonstrates that the $\ell_2$-loss function yields visibly better performance than the $\ell_1$-loss function. This pattern was consistently observed across nearly all test cases. \par
Given that the model utilized 105 2D training MRIs compared to several thousands in other frameworks (MNIST database: 60,000 training images and 10,000 test images), this reflects satisfactory performance and highlights the potential of deep learning-based models to deliver meaningful results even in scenarios where medical imaging datasets are small and less standardized. 

\begin{table}[h]
\centering %
\caption{Model errors (test data) to predict gadobutrol distribution 24 hours post-injection based on different scan intervals used as training input.}%
\begin{tabular*}{\textwidth}{@{\extracolsep\fill}lcccccc@{\extracolsep\fill}}
\toprule
 Evaluation metrics & pre-injection & 1-2 hours & 3-5 hours & 5-7 hours & 7-9 hours & 1-9 hours \\
\midrule
Mean Squared Error & \SI{7e-3}{} & \SI{2e-3}{} & \SI{6e-3}{} & \SI{5e-3}{} & \SI{4e-3}{} & \SI{1e-3}{} \\
Mean Absolute Error & \SI{3.8e-2}{} & \SI{1.8e-2}{} & \SI{3.5e-2}{} & \SI{3.5e-2}{} & \SI{3e-2}{} & \SI{1.3e-2}{} \\
\bottomrule
\end{tabular*}
\label{table:results}
\end{table}

\begin{figure}
\centering
\begin{adjustbox}{minipage=\linewidth,scale=1}
\begin{subfigure}{.16\linewidth}
  \caption{}
  \includegraphics[width=\linewidth]
  {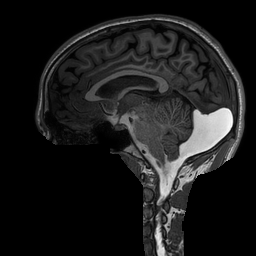}
   \label{inpt_2}
\end{subfigure}\hfill 
\begin{subfigure}{.16\linewidth}
\caption{}
   \includegraphics[width=\linewidth]
  {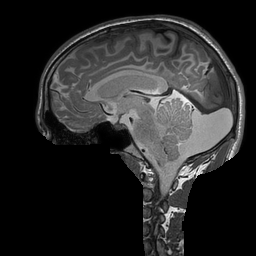}
    \label{tgt_2}
\end{subfigure}\hfill 
\begin{subfigure}{.16\linewidth}
\caption{}
  \includegraphics[width=\linewidth] {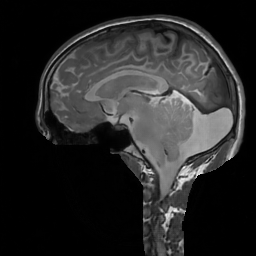}
    \label{mse_2}
\end{subfigure}\hfill 
\begin{subfigure}{.16\linewidth}
\caption{}
  \includegraphics[width=\linewidth] {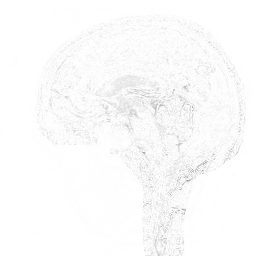}
    \label{mse_2_diff}
\end{subfigure}\hfill 
\begin{subfigure}{.16\linewidth}
\caption{}
  \includegraphics[width=\linewidth] {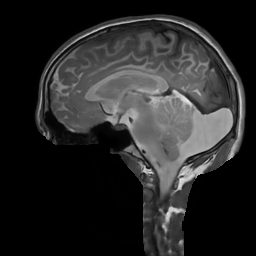}
     \label{mae_2}
\end{subfigure}
\begin{subfigure}{.16\linewidth}
\caption{}
  \includegraphics[width=\linewidth] {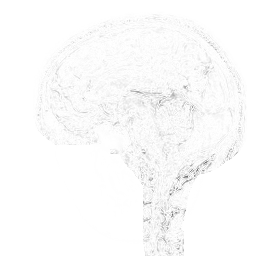}
     \label{mae_2_diff}
\end{subfigure}

\begin{subfigure}{.16\linewidth}
  \includegraphics[width=\linewidth]
  {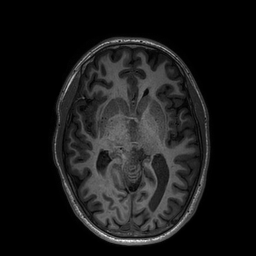}
\end{subfigure}\hfill 
\begin{subfigure}{.16\linewidth}
   \includegraphics[width=\linewidth]
  {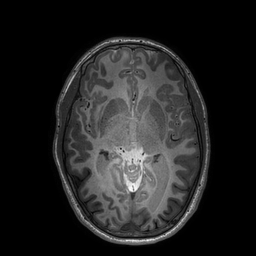}
\end{subfigure}\hfill 
\begin{subfigure}{.16\linewidth}
  \includegraphics[width=\linewidth] {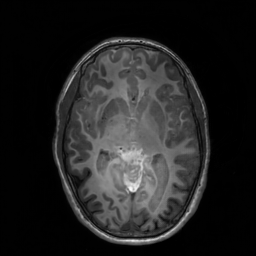}
\end{subfigure}\hfill 
\begin{subfigure}{.16\linewidth}
  \includegraphics[width=\linewidth] {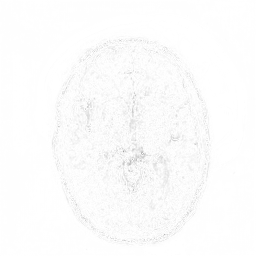}
\end{subfigure}\hfill 
\begin{subfigure}{.16\linewidth}
  \includegraphics[width=\linewidth] {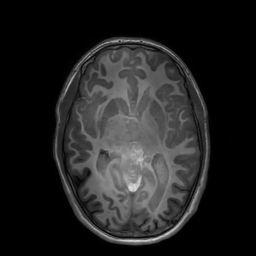}
\end{subfigure}
\begin{subfigure}{.16\linewidth}
  \includegraphics[width=\linewidth] {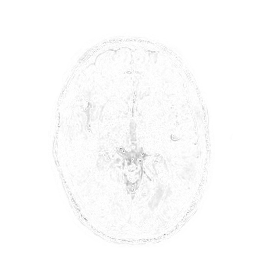}
\end{subfigure}

\vspace{1.5cm}

\begin{subfigure}{.16\linewidth}
  \includegraphics[width=\linewidth]
  {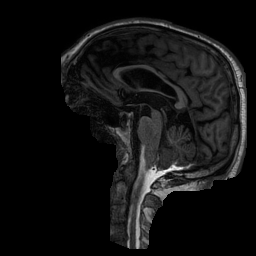}
\end{subfigure}\hfill 
\begin{subfigure}{.16\linewidth}
   \includegraphics[width=\linewidth]
  {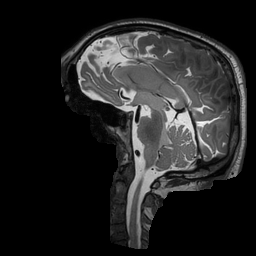}
\end{subfigure}\hfill 
\begin{subfigure}{.16\linewidth}
  \includegraphics[width=\linewidth] {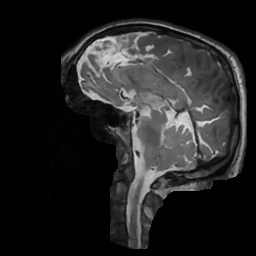}
\end{subfigure}\hfill 
\begin{subfigure}{.16\linewidth}
  \includegraphics[width=\linewidth] {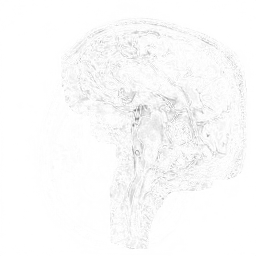}
\end{subfigure}\hfill 
\begin{subfigure}{.16\linewidth}
  \includegraphics[width=\linewidth] {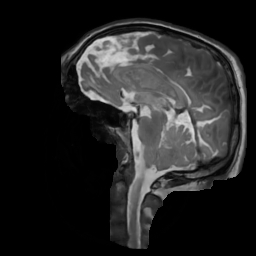}
\end{subfigure}
\begin{subfigure}{.16\linewidth}
  \includegraphics[width=\linewidth] {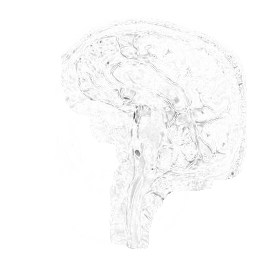}
\end{subfigure}

\vspace{0.5cm}

\begin{subfigure}{.16\linewidth}
  \includegraphics[width=\linewidth]
  {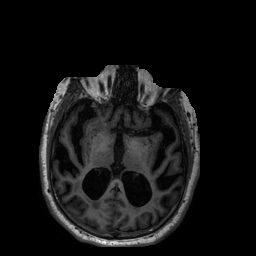}
\end{subfigure}\hfill 
\begin{subfigure}{.16\linewidth}
   \includegraphics[width=\linewidth]
  {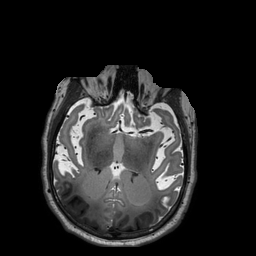}
\end{subfigure}\hfill 
\begin{subfigure}{.16\linewidth}
  \includegraphics[width=\linewidth] {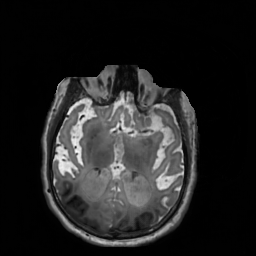}
\end{subfigure}\hfill 
\begin{subfigure}{.16\linewidth}
  \includegraphics[width=\linewidth] {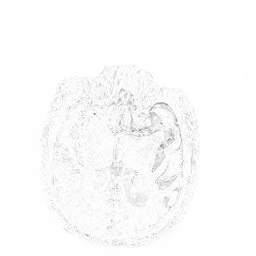}
\end{subfigure}\hfill 
\begin{subfigure}{.16\linewidth}
  \includegraphics[width=\linewidth] {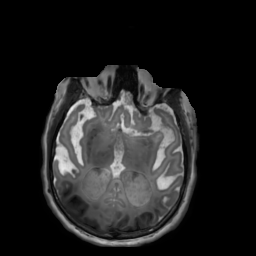}
\end{subfigure}
\begin{subfigure}{.16\linewidth}
  \includegraphics[width=\linewidth] {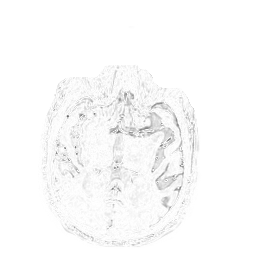}
\end{subfigure}
\end{adjustbox}
\caption{Two sample test cases (sagittal and axial planes). Rows 1 and 2 display to the first test case, while rows 3 and 4 display the second. Each column corresponds to a different set of MR images (artificial or real).  \ref{inpt_2}: real MR imaging taken within 1-2 hours post-injection. \ref{tgt_2}: real MR imaging taken approximately 24 hours after intrathecal tracer injection. \ref{mse_2}: predicted tracer distribution 24 hours post-injection ($\ell_2$ loss). \ref{mse_2_diff}: absolute difference between \ref{tgt_2} and \ref{mse_2}. \ref{mae_2}: predicted tracer distribution 24 hours post-injection ($\ell_1$loss). \ref{mae_2_diff}: absolute difference between \ref{tgt_2} and \ref{mae_2}.}
\label{fig:output2}
\end{figure}
Table \ref{table:results} summarizes achieved errors on the test data set for different MR imaging time intervals used as training data. Mean squared errors resulted in the order of magnitude \SI{1e-3}{}, while mean absolute errors were in the order of magnitude \SI{1e-2}{}. Remarkably, training solely on data obtained 1-2 hours post-injection showed performance (MSE \SI{2e-3}{}) comparable to the model trained with tracer-enriched MR scans obtained 3-5 hours, 5-7 hours, or 7-9 hours after intrathecal gadobutrol injection (MSEs \SI{6e-3}{}, \SI{5e-3}{}, \SI{4e-3}{}, respectively). Train and test losses during the U-net training process ($\ell_2$-loss) are displayed in Appendix \ref{fig:curves}. \par
Two randomly drawn test cases showing real and predicted MR imaging are displayed in Figure \ref{fig:output2}.  
The T1-weighted MRI scans illustrate two sample test cases (sagittal and axial planes) of CSF tracer (gadobutrol) distribution in the subarachnoid and intraventricular spaces, taken 1–2 hours (first column) and 24 hours (second column) after intrathecal tracer injection.
At the first time-point, enhancement with tracer (gadobutrol) induces an increased signal (white), whereas unenhanced CSF is dark. It is interesting to note that in in Patient 1, the sagittal images (upper row) at 1-2 hours after tracer injection show enhancement within an enlarged cisterna magna dorsal to the cerebellum and in the cisternal spaces ventral to the brain stem. The enhancement is well predicted by the neural network. Furthermore, there is yet no ventricular tracer reflux. At 24 hours (second column), the tracer has reached all parts of the ventricular system, though hypointense to the basal cisternal spaces (reflux grade 3). The level of enhancement is well predicted by the network. Ventricular reflux grade 3 is also seen in in Patient 2 (lower two rows), where we also note that the frontal enhancement is well-predicted. In both patients, tracer enhances unevenly in cerebral sulci at the surface. These features are all predicted well as shown in columns \ref{mse_2} and \ref{mae_2}. The involved medical experts found that using the $\ell_2$-loss generally predicts anatomical structures with more precision compared to an $\ell_1$-loss function. This is confirmed comparing columns \ref{mse_2_diff} and \ref{mae_2_diff}, which show the absolute pixel-wise error of the predictions compared to the target column \ref{tgt_2}. Differences between the two loss functions are most visible in the frontal and parietal lobe.

\subsection{Ventricular reflux grading performance compared with neuroradiologists' assessments}
The U-net model trained on scans displaying the tracer distribution within the first two hours after tracer injection was further evaluated with the ventricular reflux grading scheme. For this purpose, three human raters (board-certified neuroradiologists) were asked to assign the ventricular reflux grading on the test set of size \text{N\textsubscript{test}} = 31 comprising  sagittal-axial scan pairs that display tracer distribution after 24 hours. Grades were assigned to the original scans taken at this time (referred to as ground truth, GT) and  corresponding predicted images generated by our model. Raters were blinded both for clinical information and whether an image was real or artificial. Raters were also blinded to pre-injection imaging and evaluated only the scans obtained at the 24h mark. To avoid bias by memorizing patient-specific characteristics, rating was done in two installments such that at no time human raters had the chance to compare real and artificial scans of the same individual. Furthermore, this procedure was done for both loss functions to assess different model architecture performance.  \par
\begin{figure}
    \centering
    \begin{subfigure}[b]{0.32\textwidth}
        \centering
        \includegraphics[width=\textwidth]{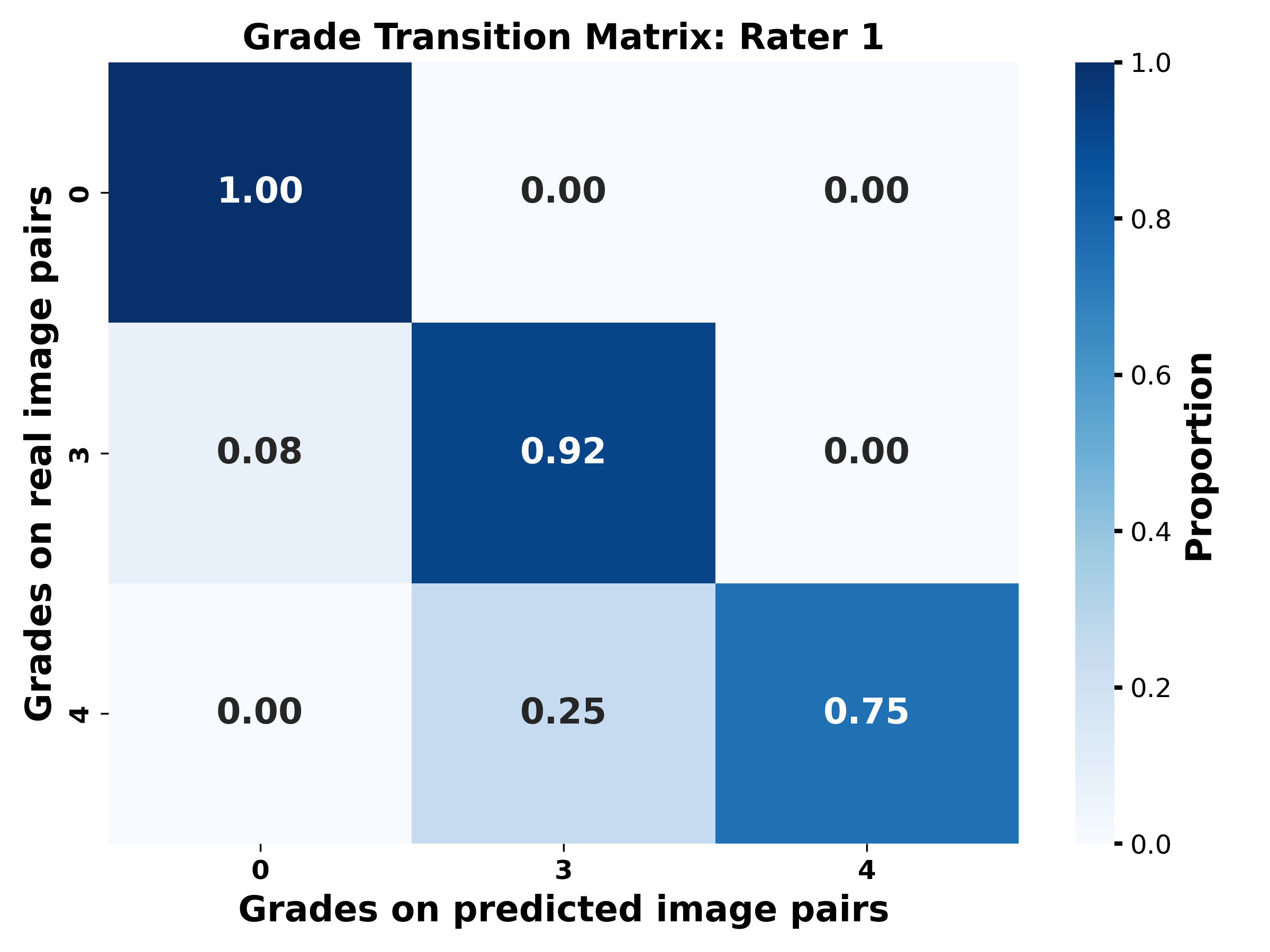}
        \caption{}
        \label{fig:heatmap_rater1}
    \end{subfigure}
    \begin{subfigure}[b]{0.32\textwidth}
        \centering
        \includegraphics[width=\textwidth]{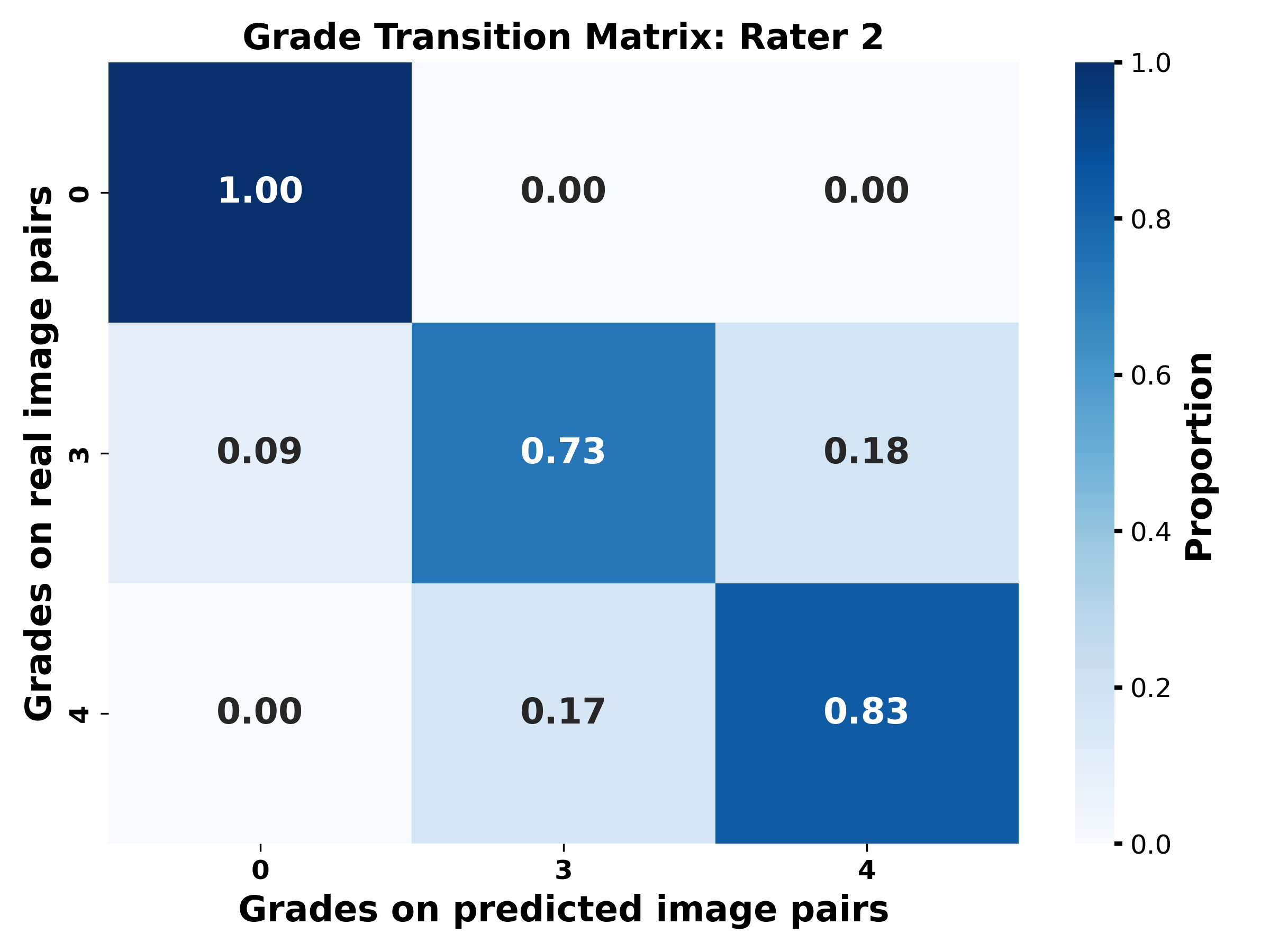}
        \caption{}
        \label{fig:heatmap_rater2}
    \end{subfigure}
    \begin{subfigure}[b]{0.32\textwidth}
        \centering
        \includegraphics[width=\textwidth]{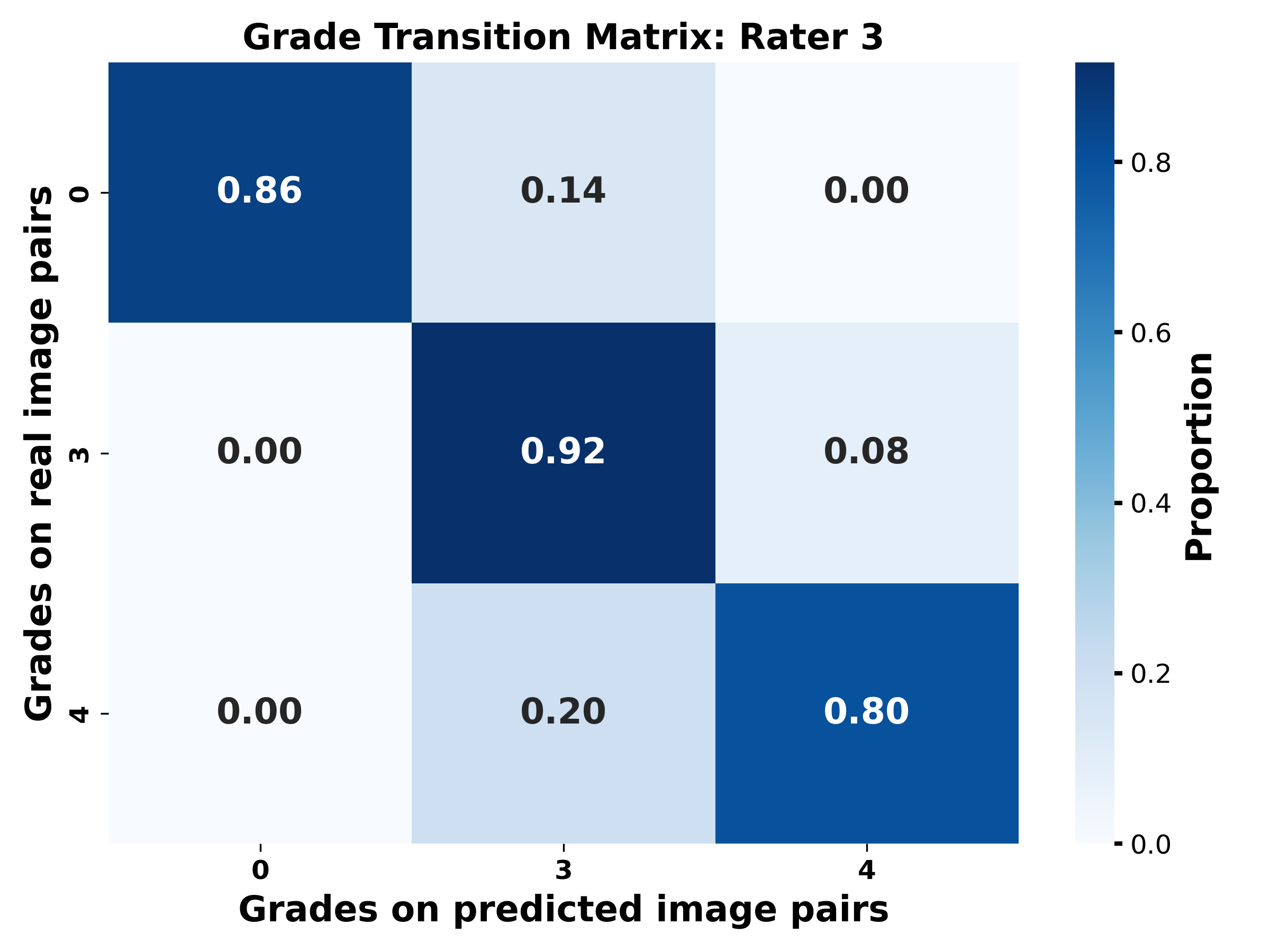}
        \caption{}
        \label{fig:heatmap_rater3}
    \end{subfigure}
    
    \caption{Grade transition matrices for Raters 1, 2, and 3. Each matrix row illustrates the percentage of grades assigned to predicted MRI slices transitioning to grades assigned to real ones, so that in each row, all entries sum up to 1.}
    \label{fig:heatmaps}
\end{figure}

Because of a natural inter-rater variability in ventricular reflux grading\cite{Eide2020}, we further used grade transition matrices (horizontal transition) to evaluate performance on the test set for each rater, shown in Figure \ref{fig:heatmaps}. These matrices illustrate the alignment between grades assigned to real axial-sagittal image pairs per subject and their corresponding pairs of model predictions. For instance, Figure \ref{fig:heatmap_rater1} shows that 92\% of individuals in the test set who were assigned a grade of 3 by Rater 1 on the real images also received the same grade 3 from Rater 1 on the predicted images. \par 
Rater 1 exhibited a strong agreement, with most cases aligning along the diagonal, indicating consistency in grading. However, slight discrepancies were observed in lower-grade assignments, where some predicted grades deviated by one category. Rater 2 showed a similar pattern, though with a marginally higher proportion of off-diagonal transitions, particularly for grade 3 cases being over-predicted as grade 4 or grade 4 cases being under-predicted as grade 3. Noticeably, none of the scans that were assigned grade 0 on the real MRI were assigend grade 4 on the artificial MRI for any of the raters, likewise the other way around. \par

Lastly, Rater 3 demonstrated the highest variability, with 14\% of grade 0 cases being assigned grade 3 on the predictions, which provides a contrast to Raters 1 and 2. This behavior could be explained by the fact that the distinction between grade 0 and 3 hinges solely on the presence or absence of signal  enhancement at 24h, making the classification potentially ambiguous in regions of interest with low enhancement.  \par
In summary, the findings of this work demonstrate that the model’s predictions lead to consistent grades similar to those that are obtained with the ground truth. Variations observed in borderline cases are likely to reflect differences in individual grading tendencies rather than significant limitations of the model itself. It is worth to point out thtat grades 1 and 2 do not occur because the data used for grading did not exhibit transient enrichment of the lateral ventricles, which is the defining characteristic of these grades. Tracer distribution patterns observed after 24 hours generally corresponded to Grades 3 or 4, where the enhancement is either lasting but less intense than the surrounding subarachnoid spaces (Grade 3) or fully isointense (Grade 4).

\section{Discussion}\label{discussion}
The main observation of this study was that U‐net based deep learning provided a high pixel‐wise accuracy in predicting the 24‐hour distribution utilizing images obtained within 1‐2 hours post‐injection. These results suggest that the current clinical practice of extended MR imaging sessions to capture later stages of tracer distribution could be reduced without sacrificing diagnostic accuracy. This may provide for improved cost‐efficiency and patient throughput. \par

Traditionally, phase‐contrast magnetic resonance imaging (PC-MRI) has been used to assess CSF flow in vivo\cite{Eide2021_2}; however, that method has limitations by its ability to only image water motions at short term and over short distances, and tracer studies using contrast agents as CSF tracers may have advantages. In particular, traditional PC-MRI only measures back-and-forth movement on the second scale and includes neither netflow nor the slow pulsations of vasomotion, which may be important for solute transfer. Furthermore, the methodology is not well-suited for the thin and complex spaces of the subarachnoid space~\cite{Rivera2024}. Intrathecal contrast enhanced MRI, such as those described in~\cite{Ringstad2017,Ringstad2018}, provide a unique view into the slow transport (over days) of solutes within the CSF and the brain. In fact, as shown by \cite{Eide2021_2}, neither ICP score nor aquaductal flow, as measured with PC-MR directly, correlates with contrast enhancement. As such, it is remarkable that the U-net approach, learning from just 2D sample images, manages to predict then enhancement pattern the next day to a reasonable accuracy presented above. 

A disadvantage with tracer studies is the need for several MRI acquisitions, which not only place significant burden on patients, but are also resource-intensive.  \par

In addition to error metrics common in neural network regression, we evaluated the model performance by comparing the grading of ventricular reflux on both real scans and those predicted by our model. Inter-rater reliability between the human raters and the model was generally higher than the reliability between human raters alone, indicating that the  artificially generated MR scans of the human brain could be equally conclusive for diagnostic CSF assessment. \par

While the proposed U-net-based deep learning model demonstrates potential in predicting CSF tracer distribution and assessing ventricular reflux, two limitations of this study should be noted. First, our study is based on a small cohort size of 136 patients with different underlying diseases, which may limit the generalizability of the model's performance across diverse populations. Second, we utilize a 2D U-net architecture for image prediction, which limits its effectiveness in capturing the full complexity of 3D anatomical structures. Utilization of 3D volumetric data is desirable in the clinical setting, as neuroradiologists commonly assess CSF dynamics and ventricular reflux grading scores across multiple planes. \par

Future prospects are the use of sufficient computational resources that will give further research opportunities, ideally leveraging volumetric 3D data and integrating this data into a 3D U-net architecture. Another promising direction for future research lies in predicting specifically the tracer clearance process, particularly after the peak distribution at roughly 24 hours post-injection. It remains an open question whether data from the first one to ten hours post-injection are satisfactory to predict decline of tracer signal with a similar deep-learning approach, which would provide further insights into potentially disturbed flow patterns. Understanding the clearance process is crucial, as delays in this process serve as a pathophysiological indicator of several conditions related to neurodegenerative decline, such as in idiopathic normal pressure hydrocephalus (iNPH), Alzheimer's, and Parkinson's disease \cite{Lopes2022}. Furthermore, integrating clinical measurements, outcomes and longitudinal data, such as patient demographic information, neurocognitive assessment scores, blood work, CSF biomarkers, and patient-specific tracer concentration could enhance the prediction performance and offer more comprehensive evaluations of the model's predictive capabilities and clinical utility. \par

In summary, U‐net based deep learning was found useful for predicting tracer distribution 24 hours after its intrathecal administration, providing the network with 2D MRI data. This could have clinical implications by reducing the number of MRI acquisitions; however, further research is needed to implement this methodology for clinical use.

\section*{Data availability statement}
The data that support the findings of this study are available from the corresponding author upon reasonable request.

\section*{Author contributions}
MR conducted the experiments, performed analysis, and drafted the manuscript under supervision of FH, BW, KAM, and PKE. LMV prepared the experimental data. GR, BW, and OL provided grading of medical images. All authors have read the manuscript, provided critical feedback and approved the final version.

\bibliographystyle{unsrt}  
\bibliography{CSFMLPaper}  

\appendix

\section{Loss curves for U-net training and testing over 250 epochs ($\ell_2$ loss)}

\begin{figure}[hbt]
\centering
\begin{adjustbox}{minipage=\linewidth,scale=0.95}
\begin{subfigure}{.32\linewidth}
  \caption{}
  \includegraphics[width=\linewidth]{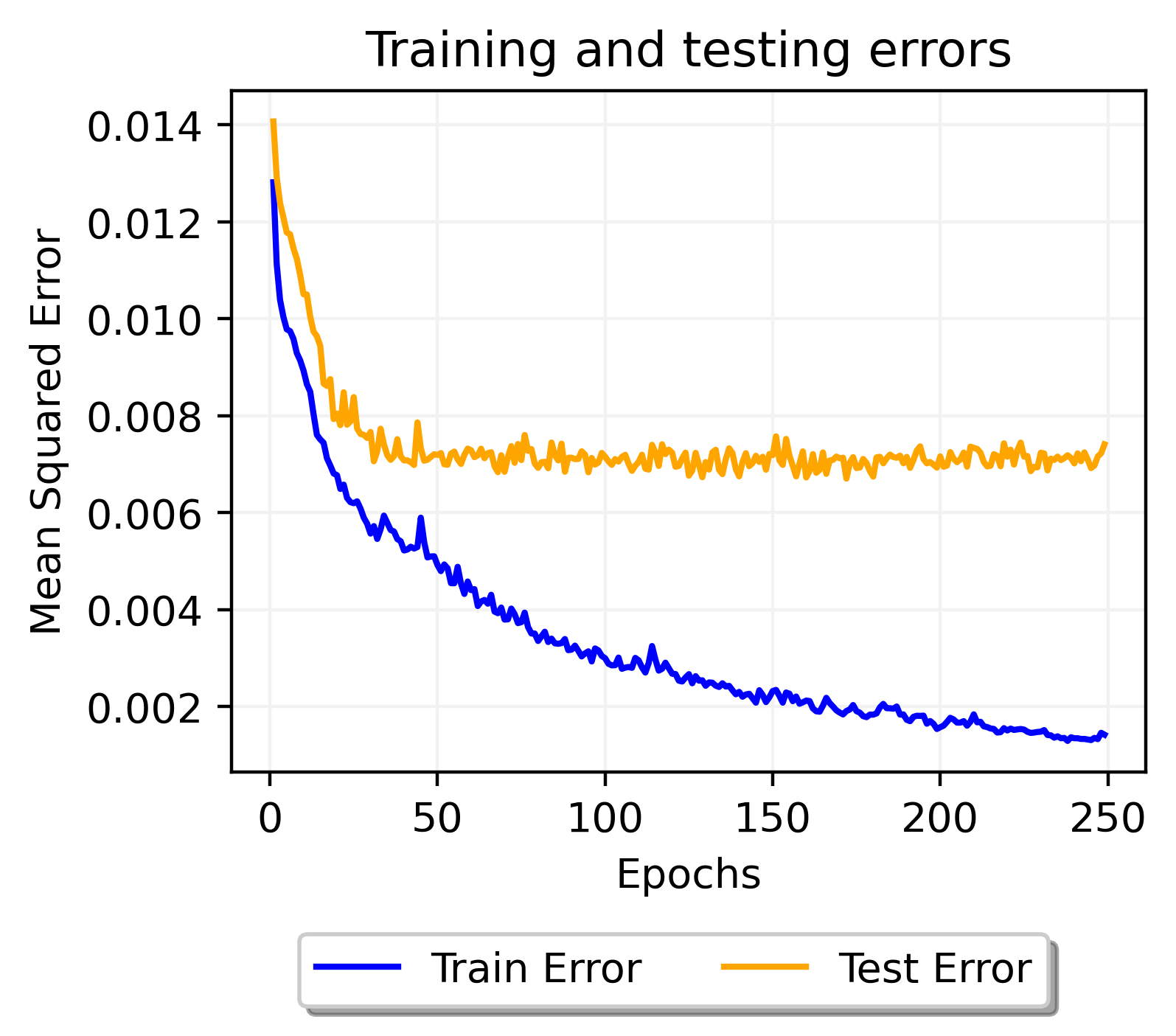}
  \label{fig:curve1}
\end{subfigure}\hfill 
\begin{subfigure}{.32\linewidth}
\caption{}
  \includegraphics[width=\linewidth]{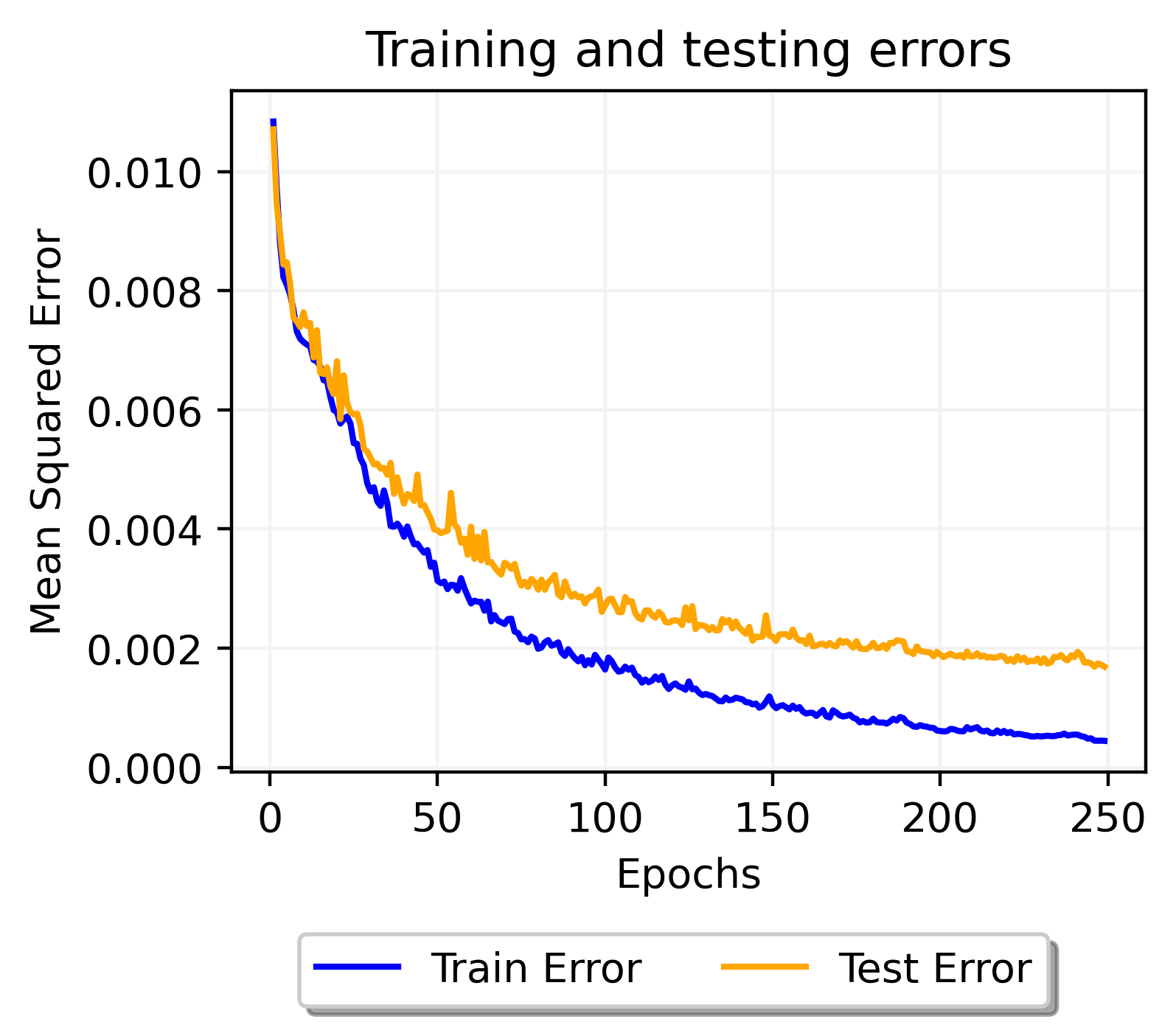}
  \label{fig:curve2}
\end{subfigure}\hfill 
\begin{subfigure}{.32\linewidth}
\caption{}
  \includegraphics[width=\linewidth]{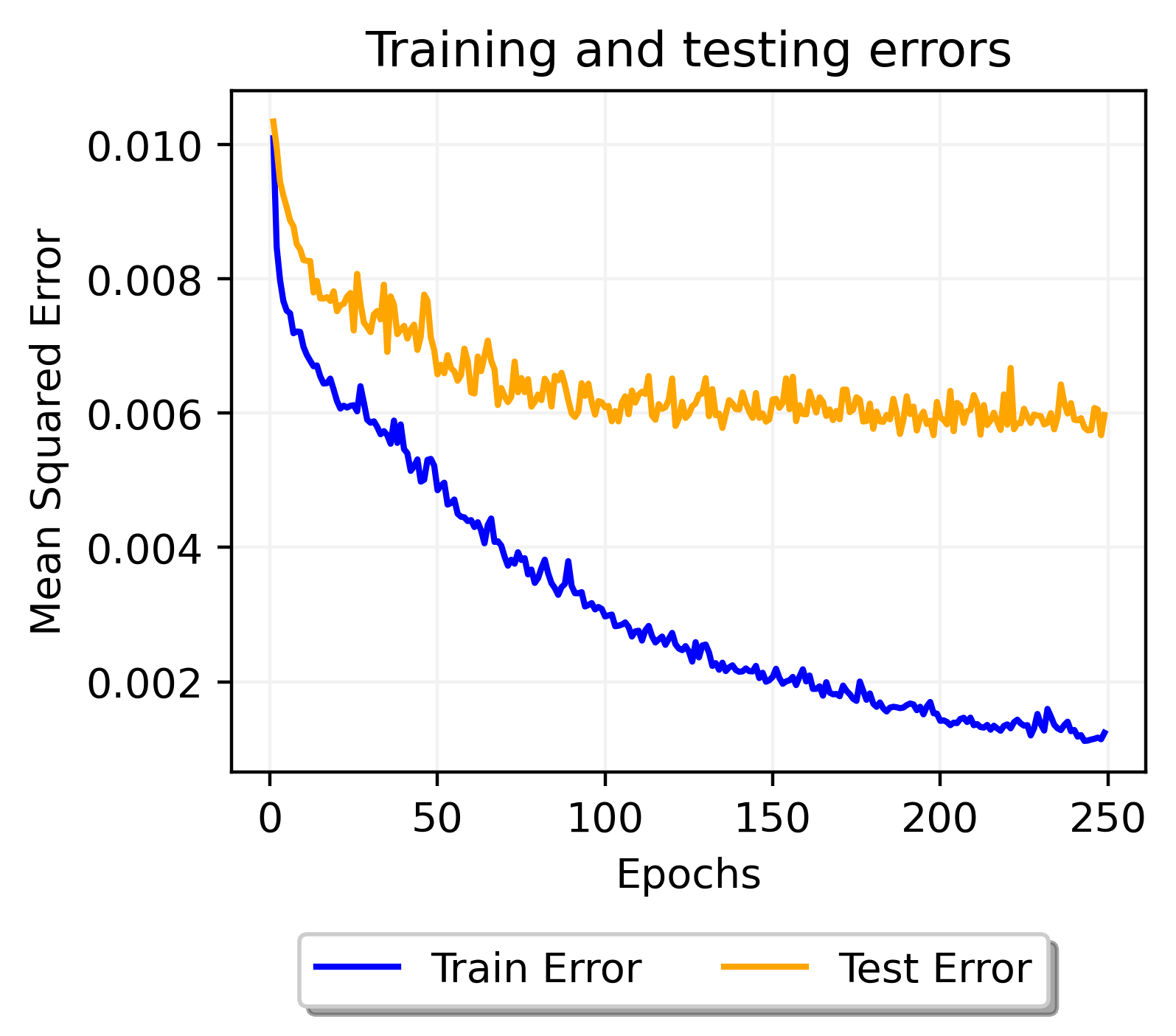}
    \label{fig:curve3}
\end{subfigure}
 
\begin{subfigure}{.32\linewidth}
\caption{}
  \includegraphics[width=\linewidth]{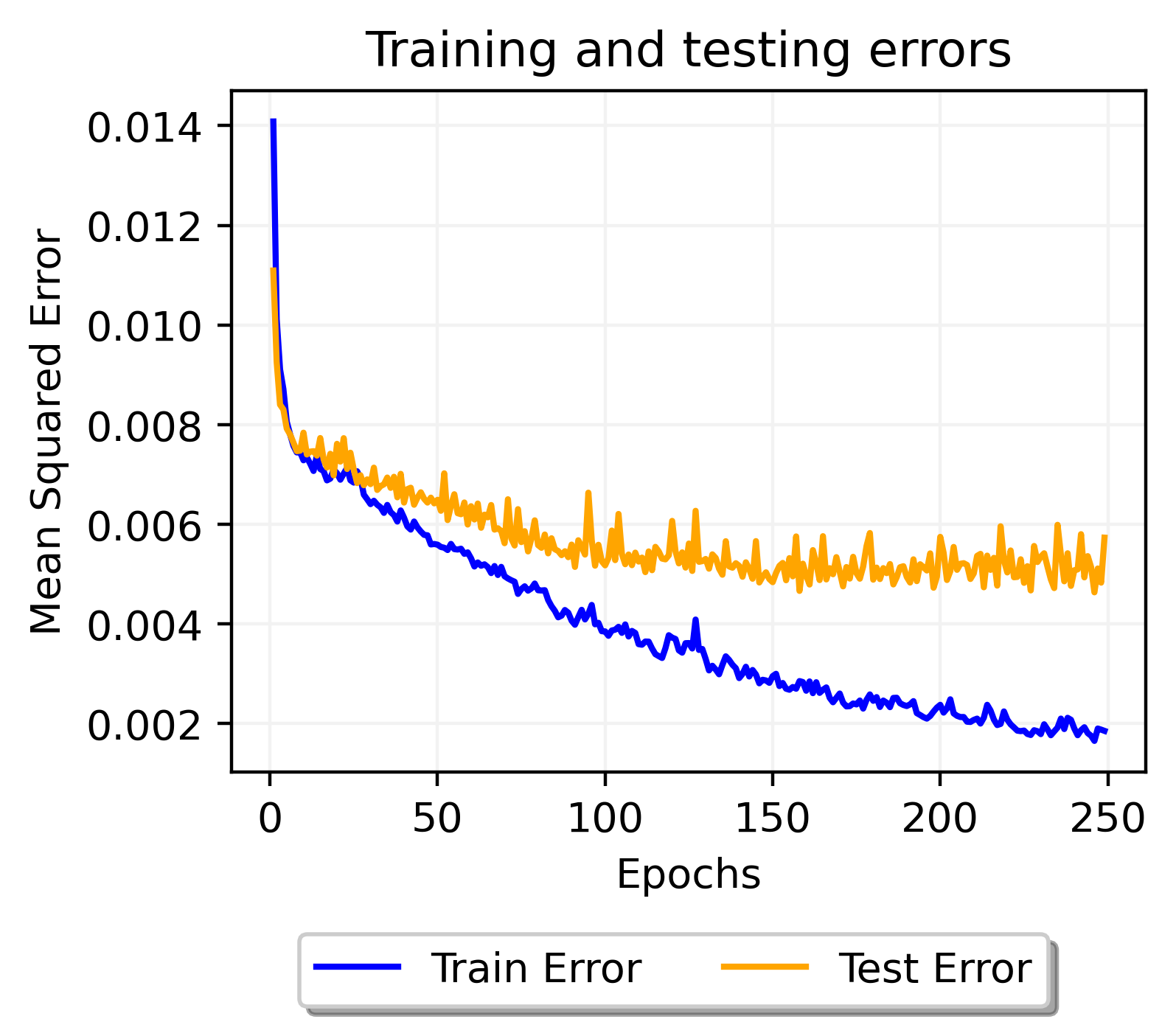}
    \label{fig:curve4}
\end{subfigure}\hfill 
\begin{subfigure}{.32\linewidth}
\caption{}
  \includegraphics[width=\linewidth]{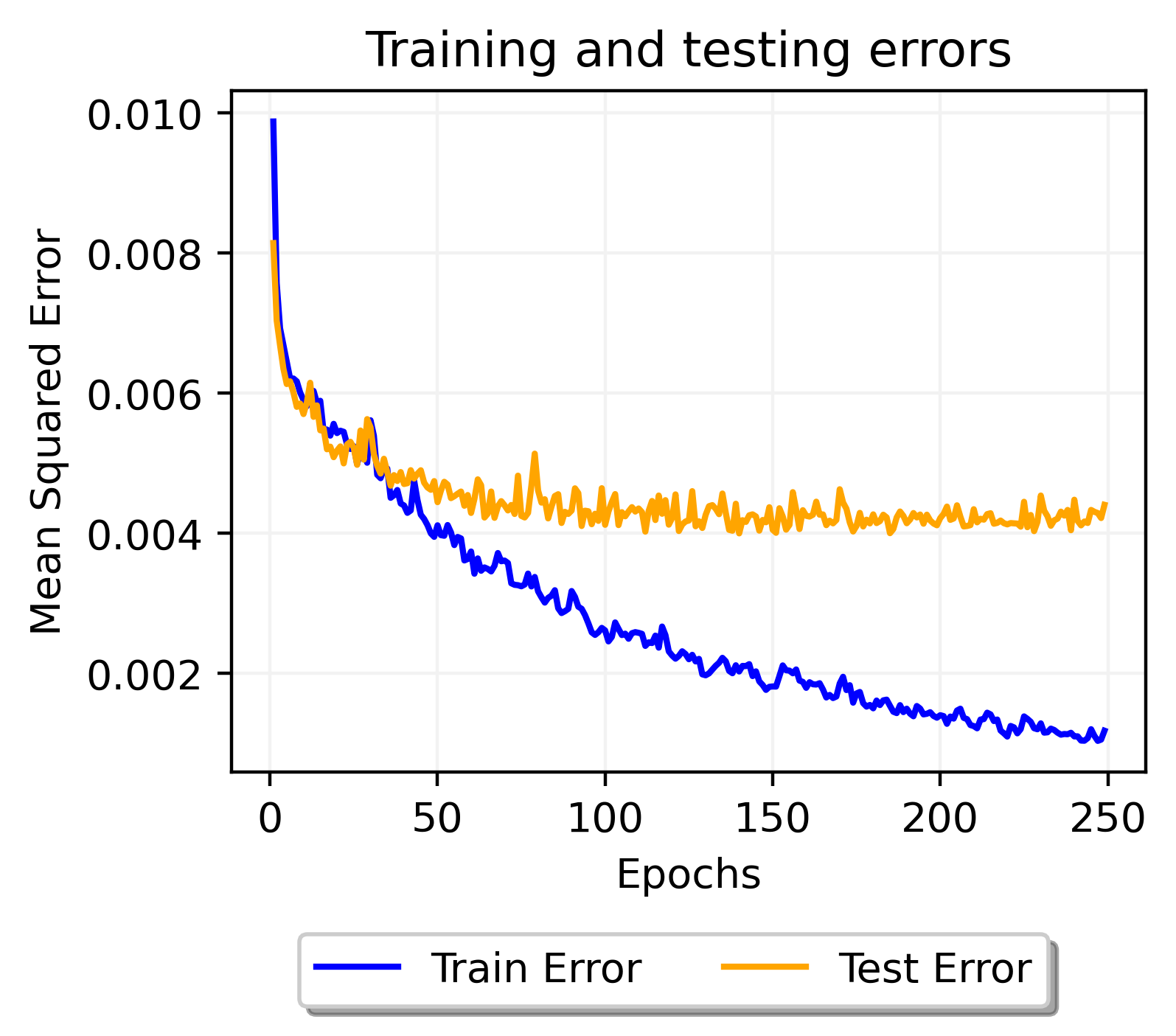}
    \label{fig:curve5}
\end{subfigure}\hfill 
\begin{subfigure}{.32\linewidth}
\caption{}
  \includegraphics[width=\linewidth]{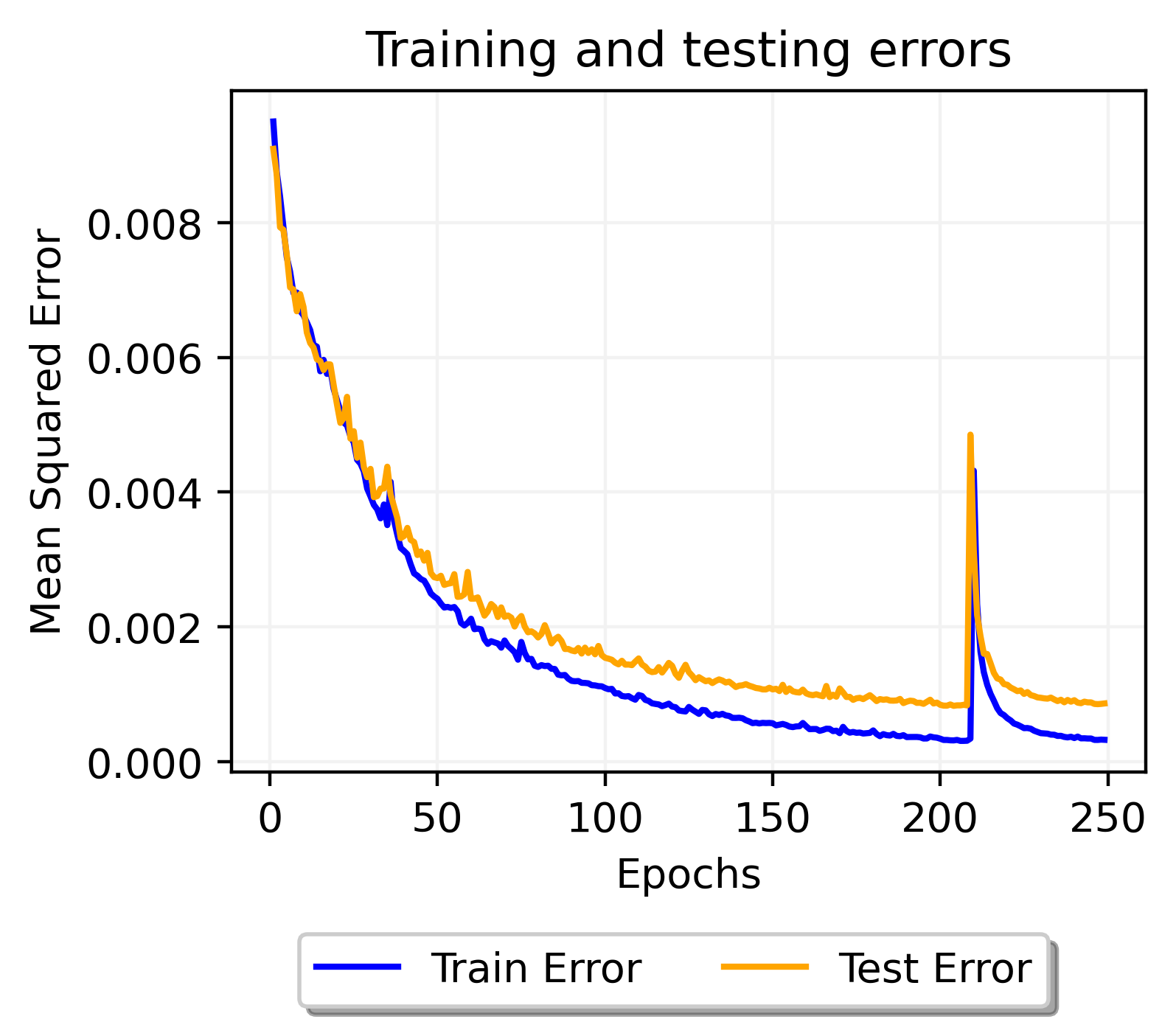}
    \label{fig:curve7}
\end{subfigure}
\end{adjustbox}
\caption{\ref{fig:curve1}: Prediction from baseline (pre-injection). \ref{fig:curve2}: Prediction from 1-2 hours post-injection. \ref{fig:curve3}: Prediction from 3-5 hours post-injection. \ref{fig:curve4}: Prediction from 5-7 hours post-injection. \ref{fig:curve5}: Prediction from 7-9 hours post-injection. \ref{fig:curve7}: Prediction from 1-9 hours post-injection.}
\label{fig:curves}
\end{figure}

In the loss curve on Figures \ref{fig:curve1}, \ref{fig:curve3}, \ref{fig:curve4} and \ref{fig:curve5}, we can see that the curve of loss tends to flatten as the epochs number increases, indicating that the training model converges. In \ref{fig:curve2} and \ref{fig:curve7}, the gap between training and testing error is smaller, indicating less overfitting and better model performance, which is in accordance to the evaluation metrics shown in Figure \ref{table:results}.

\end{document}